\documentclass[twocolumn,times]{aastex631} 
\usepackage{graphicx}
\usepackage{mathtools} 
\usepackage{natbib}
\usepackage{amsmath}
\usepackage{verbatim}
\usepackage{appendix}
\usepackage{multirow,tabularx}
\usepackage{outlines}
\usepackage[version=4]{mhchem} 
\usepackage{graphicx}
\usepackage{comment}
\usepackage{float}
\usepackage{colortbl}
\usepackage{xkcdcolors}

\newcommand\Teff{$T_\mathrm{eff}$}
\newcommand\Tten{$T_{10}$}
\newcommand\Msun{$M_\mathrm{\odot}$}
\newcommand\Mjup{$M_\mathrm{J}$}

\defcitealias{Morely:2024_SonoraDiamondback}{M24}
\defcitealias{Iyer:2023_SphinxMdwarfAtm}{I23}
\defcitealias{Chabrier:2023_CD21newHBMM}{C23}
\newcommand\SPX{\citetalias{Iyer:2023_SphinxMdwarfAtm}}
\newcommand\DBK{\citetalias{Morely:2024_SonoraDiamondback}}
\newcommand\CBPD{\citetalias{Chabrier:2023_CD21newHBMM}}

\shortauthors{Davis et al.}

\begin{document}

\title{The Sonora Substellar Atmosphere Models VI. Red Diamondback: \\Extending Diamondback with SPHINX for Brown Dwarf Early Evolution}

\correspondingauthor{C.  Evan Davis}
\email{c.evan.davis@ucsc.edu}

\author[0009-0005-3386-6091]{C.  Evan Davis}
\affiliation{Department of Astronomy and Astrophysics, University of California, Santa Cruz, 1156 High St, Santa Cruz, CA 95064, USA}

\author[0000-0002-9843-4354]{Jonathan J.  Fortney}
\affiliation{Department of Astronomy and Astrophysics, University of California, Santa Cruz, 1156 High St, Santa Cruz, CA 95064, USA}

\author[0000-0003-0971-1709]{Aishwarya Iyer}
\altaffiliation{NASA Postdoctoral Program Fellow}
\affiliation{Goddard Space Flight Center, Greenbelt, MD 20771, USA}

\author[0000-0003-1622-1302]{Sagnick Mukherjee}
\affiliation{Department of Astronomy and Astrophysics, University of California, Santa Cruz, 1156 High St, Santa Cruz, CA 95064, USA}
\affiliation{School of Earth and Space Exploration, Arizona State University, Tempe, AZ, USA \\ }

\author[0000-0002-4404-0456]{Caroline V. Morley}
\affiliation{Department of Astronomy, University of Texas at Austin, 2515 Speedway, Austin, TX 78712, USA}

\author[0000-0002-5251-2943]{Mark S. Marley}
\affiliation{Lunar and Planetary Laboratory, University of Arizona, Tucson, AZ 85721, USA}

\author[0000-0001-6247-8323]{Michael Line}
\affiliation{School of Earth and Space Exploration, Arizona State University, 525 E.  University Dr., Tempe, AZ 85281}

\author[0000-0002-0638-8822]{Philip S. Muirhead}
\affiliation{Department of Astronomy and Institute for Astrophysical Research, Boston University, 725 Commonwealth Ave., Boston, MA 02215, USA}

\begin{abstract}
We extend the Sonora Diamondback brown dwarf evolution models to higher effective temperatures to treat the evolution of younger, higher mass objects.  Due to an upper temperature limit of \Teff$=$2400 K in the original Sonora Diamondback model grid, high mass objects ($M\geq$ 0.05 \Msun{} $=$ 52.4 \Mjup) were limited to ages of $\gtrsim$ 100 Myr.  To include the early evolution of brown dwarfs at \Teff{} $>$ 2400 K, we use existing and new SPHINX cloud-free model atmosphere calculations of temperature structures of M-type atmospheres.  These atmospheres range from \Teff{} 2000--4000 K, log($g$) 3.0--5.5, and metallicity [M/H] $-$0.5 to $+$0.5.  This combination of Diamondback and SPHINX atmospheres, with a transition across \Teff{} 2000--2400 K, allows us to calculate evolution tracks, and infrared photometry and colors, for ages $>$ 1 Myr and masses from above the hydrogen burning minimum mass down to planetary masses.  The Hayashi phase of massive brown dwarf evolution (ages $<$ 10--100 Myr) at low surface gravity leads to nearly constant \Teff{} values, at effective temperatures much lower than would be obtained from simply extrapolating backwards from evolution tracks at older ages.


\end{abstract}


\section{Introduction} \label{sec:intro}

\begin{table*}[]
    \centering
    \caption{Model Grid Parameters}
        \begin{tabular}{lccc}
        \hline\hline 
        Parameter {[}unit{]}       & Sonora ``Red Diamondback''      & Sonora ``Diamondback''               & SPHINX                           \\ \hline
        Reference                  & This Work                       & \cite{Morely:2024_SonoraDiamondback} & \cite{Iyer:2023_SphinxMdwarfAtm} \\
        $\mathrm{T_{eff}}$ {[}K{]} & 900--4000, $\Delta$100          & 900--2400, $\Delta$100               & 2000--4000, $\Delta$100          \\
        log($g$ {[}cm/s$^2${]})    & 3.0--5.5, $\Delta$0.25          & 3.5--5.5, $\Delta$0.5                & 4.0--5.5, $\Delta$0.25           \\
        {[}M/H{]} {[}dex{]}        & $-$0.5--$+$0.5, $\Delta$0.5     & $-$0.5--$+$0.5, $\Delta$0.5          & $-$1.0--$+$1.0, $\Delta$0.25     \\
        C/O                        & 0.5 (\SPX), 0.458 (\DBK)        & 0.458                                & 0.3--0.9, $\Delta$0.2            \\ 
        $f_{\mathrm{sed}}$         & 2 (\DBK{} ``hybrid'' only)      & 1,2,3,4,8                            & ---                              \\ \hline
        \end{tabular}
    \label{tab:SonoraModelComparisons}
\end{table*}

Three decades after the first brown dwarf was discovered \citep[GJ229B,][]{Nakajima:1995_GJ229B}, many hundreds of these substellar objects have been found in the solar neighborhood \citep{Kirkpatrick:2021_BDSurvey}.  As brown dwarfs are quite faint even in their brightest wavelengths (the near- or mid-infrared), much of the observed brown dwarf population is comprised of very nearby objects, which are the older ``field'' brown dwarfs.  However, a growing sample of young brown dwarfs have been identified in several nearby young associations \citep{Gagne:2014_BANYAN2} and young star-forming regions and clusters like NGC1333 \citep{Langeveld:2024_JWSTyoungBDsurvey} and the Orion Nebula Cluster \citep{Luhman:2024_ONCBDs}.  Additionally, JWST's ability to peer through the dust and gas that often surrounds these objects in their infancy has allowed the community to begin characterizing these objects at very young ages.  For instance, \citet{Manjavacas:2024_YoungBDwJWST} obtained exquisite R$\sim$2700 spectra of TWA 28 and TWA 27A, both late M-type brown dwarfs in the $\sim$10 Myr-old TW Hydrae Association.

Characterizing the population of young brown dwarfs would illuminate the formation mechanisms that operate in the mass range between planets and stars, but this effort faces several challenges.  To what degree brown dwarfs form like planets via disk-fragmentation, or like low-mass stars via fragmentation of giant molecular clouds, is still an active question \citep{Palau:2024_YoungBDformation}.  Which formation process is more efficient and where the two processes overlap is thought to be dependent on mass.  However, the vast majority of these objects are free-floating, making direct measurements of mass difficult if not impossible.  Additionally, because brown dwarfs never reach a main sequence and continuously cool over time, a strong degeneracy between mass and age exists: a small brown dwarf that was recently born and has only just started to cool can be the same effective temperature and of a similar spectral type as a massive brown dwarf that has had more time to cool \citep{Burrows:1989_BDEvo}.  Thus, while photometric and spectral measurements of these objects give us valuable insight into their fundamental properties, their formation histories are difficult to constrain without an additional component that breaks this mass-age degeneracy.



Thermal evolution models bridge this gap.  These models simulate how the fundamental properties (luminosity, radius, effective temperature, surface gravity, etc) of an object of a given mass evolve.  With an empirical luminosity (or brightness in a bandpass) and estimated age of an observed brown dwarf in hand, one can compare to theoretical luminosities from evolution tracks to constrain the observed object's mass.  An early calculation of such model grids was \citet{Burrows:1989_BDEvo}, with subsequent studies including non-grey atmospheric boundary conditions \citep{Burrows:1997_NonGrayBDEvo, Baraffe:2003_CoolBDEvo}, updated molecular opacities \citep{Phillips:2020_ATMO2020evo, Marely:2021_SonoraBobcat}, and modeling of cloud condensation and sedimentation processes \citep{Chabrier2000_DustyBDEvo,Saumon&Marley:2008_LTdwarfEvo,Morely:2024_SonoraDiamondback}.

Objects with substellar masses can readily traverse multiple spectral classes, across M, L, T, and Y.  Therefore, the model atmospheres that serve as the upper boundary condition for evolution models must be computed over a very wide range of effective temperatures (\Teff) and surface gravities (log($g$)).  For objects older than $\sim$100 Myr, \Teff$\leq$2400 K and log($g$)$\geq$4 are typical of the population.  It is these lower \Teff{} conditions where we have focused the recent \emph{Sonora} set of models \citep{Marely:2021_SonoraBobcat,Karalidi:2021_SonoraCholla,Morely:2024_SonoraDiamondback,Mukherjee:2024_SonoraElfOwl}, and in previous evolutionary calculations \citep{Saumon&Marley:2008_LTdwarfEvo}.  Higher \Teff{} values require atmospheric opacities and chemistry appropriate for M-type dwarfs, which is outside of the bounds of Sonora atmospheres.  Our aforementioned evolution models therefore needed to extrapolate atmospheric boundary conditions at higher \Teff{} values, which we will examine in this paper.

Other groups have recently modeled brown dwarf evolution across a full range of masses and ages, but sometimes with simplifications to the upper boundary condition \citep[e.g.][with cloud-free atmospheres that also lacked some high-temperature gaseous opacities]{Chabrier:2023_CD21newHBMM}.  The early evolution of brown dwarfs is an inherently interesting area where uncertainties in the energetics of formation, energy transport, and atmospheric boundary conditions are stress-tested \citep{Baraffe:2002_EarlyBDEvo}.

To address the important need for brown dwarf thermal evolution models at young ages, we have built upon our recent grid of atmosphere and evolution models, \texttt{Sonora Diamondback} \citep[][hereafter \DBK]{Morely:2024_SonoraDiamondback}, which treated brown dwarfs at \Teff$\leq$2400 K using an upgraded version of the thermal evolution code originally presented in \citet{Throgren:2016_GiantPlanetMZ}.  Here, we use a large set of new model atmosphere temperature structures and spectra calculated using the \texttt{SPHINX} M-dwarf Spectral Grid \citep[][hereafter \SPX]{Iyer:2023_SphinxMdwarfAtm} modeling framework.  These models, with updated molecular and atomic opacities, were recently shown to compare favorably to spectra of benchmark M dwarfs, up to \Teff{}$=$4000 K.  In this work, we use cloud-free \texttt{SPHINX} upper boundary conditions above 2000 K, and our existing \texttt{Sonora Diamondback} boundary conditions for $\leq$ 2400 K (which include clouds, where appropriate), with a transition region between the two.  This allows us to calculate the evolutionary history for objects just above the hydrogen-burning minimum mass (HBMM) to Jovian masses, starting at an age of 1 Myr.



\section{Methods}


\subsection{Atmospheres} \label{sec:methods:atm}
For fully convective objects like brown dwarfs, the radiative-convective atmospheric boundary condition is critical to controlling the cooling of the interior \citep{Hubbard:1977_JovianCoolingRates, Burrows:1997_NonGrayBDEvo}.  Such a boundary connects the specific entropy of the interior and surface gravity with the \Teff{} of the atmosphere.

In this work, we use two atmosphere libraries in concert to calculate our boundary conditions: the \texttt{Sonora Diamondback} ``hybrid'' atmospheres \citepalias[see][for a full description]{Morely:2024_SonoraDiamondback} at \Teff{}$\leq$2400 K, and the recent \texttt{SPHINX} M-dwarf atmosphere grid \citepalias{Iyer:2023_SphinxMdwarfAtm} at \Teff{}$\geq$2000 K.  \texttt{SPHINX} was designed to generate spectra, pressure-temperature profiles, and abundance profiles across the M-dwarf temperature range (\Teff{}$=$2000--4000 K).  \texttt{SPHINX} uses a radiative convective thermochemical equilibrium model based on ScCHIMERA \citep{Arcangeli:2018_WASP18bAtmopshere, Bonnefoy:2018_GJ504, Mansfield:2021_HotJupDiversity} and an atmosphere climate model that has been validated against robust brown dwarf models \citep{Saumon&Marley:2008_LTdwarfEvo}. In addition to molecular lists appropriate for cool stellar atmospheres, \texttt{SPHINX} includes a range of atomic opacities (neutral: Na, K, Fe I, Mg I, Ca I, C I, Si I, Ti I, O I, ionized atoms: Fe II, Mg II , Ti II , C II , Ca II, see \SPX{} for full description), some of which are not present in the \texttt{Sonora Diamondback} models.  Both \texttt{Sonora Diamondback} and \texttt{SPHINX} iteratively solve for the pressure-dependent temperature, volume mixing ratios, condensate properties, and disk-integrated spectrum of the object's atmosphere, for models in radiative-convective equilibrium.  Lastly, to compute the convective flux in the radiative transfer calculations, \texttt{SPHINX} uses mixing length theory \citep{BohmVitense:1958_MLT} while \texttt{Sonora Diamondback} used adiabatic adjustment. Thus, only \texttt{SPHINX} can treat the possibility of inefficient super-adiabatic convection in low gravity atmospheres.

Both sets of atmospheres are pressure--temperature ($P$--$T$) profiles gridded across effective temperature, surface gravity, metallicity, and carbon-to-oxygen ratio, the values of which which are shown in Table \ref{tab:SonoraModelComparisons}.  The \DBK{} atmospheres were also gridded across $f_{\mathrm{sed}}$, the cloud sedimentation efficiency parameter. Where this work uses the \DBK{} atmospheres, we use the published value $f_{\mathrm{sed}}=2$.  In addition to the surface gravity values computed in the original \SPX{} paper, we compute additional models for surface gravities down to 3.0 in 0.25 steps for effective temperatures of 2400--3300 K and [M/H] of $-0.5$, $+0.0$, and $+0.5$.  This extension to the \SPX{} grid allows the simulation our objects back to ages of 1 Myr.

\begin{figure*}[ht]
    \centering
    \includegraphics[width=0.75\linewidth]{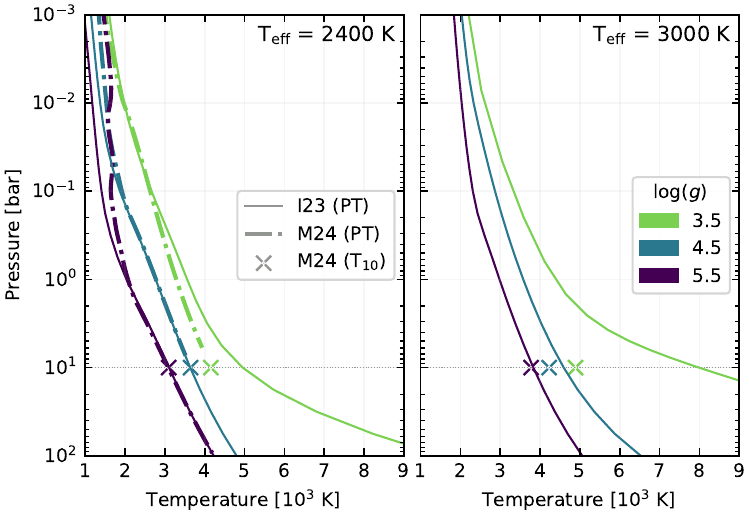}
    \caption{Selected $P$--$T$ profiles from \SPX{} (thin solid line) and \DBK{} (thick dot-dashed line) for two effective temperatures and three surface gravities.  The left panel shows atmospheres with \Teff{} $=$ 2400 K, while the right panel shows those with \Teff{} $=$ 3000 K.  We show the interpolated/extrapolated \Tten{} values from \DBK{} along the $P=$ 10 bar isobar (horizontal grey dotted line).  At 2400 K, the \DBK{} atmospheres are hotter at lower pressures than the \SPX{} atmospheres due to the inclusion of clouds that add opacity, although this has minimal impact on \Tten{}.  The \SPX{} atmospheres start to become hotter at log($g$) $\lesssim$ 4 due to the inclusion of a range of atomic opacities not found in \DBK{}.  By \Teff{} $=$ 3000 K, the \SPX{} atmospheres are hotter at each surface gravity than the extrapolated \Tten{} values from \DBK, with the discrepancy widening dramatically as one goes to lower surface gravities ($\gtrsim$ 3000 K at log($g$) $=$ 3.5).}
    \label{fig:PT_Comparison}
\end{figure*}

\begin{figure*}[ht]
    \centering
    \includegraphics[width=0.75\linewidth]{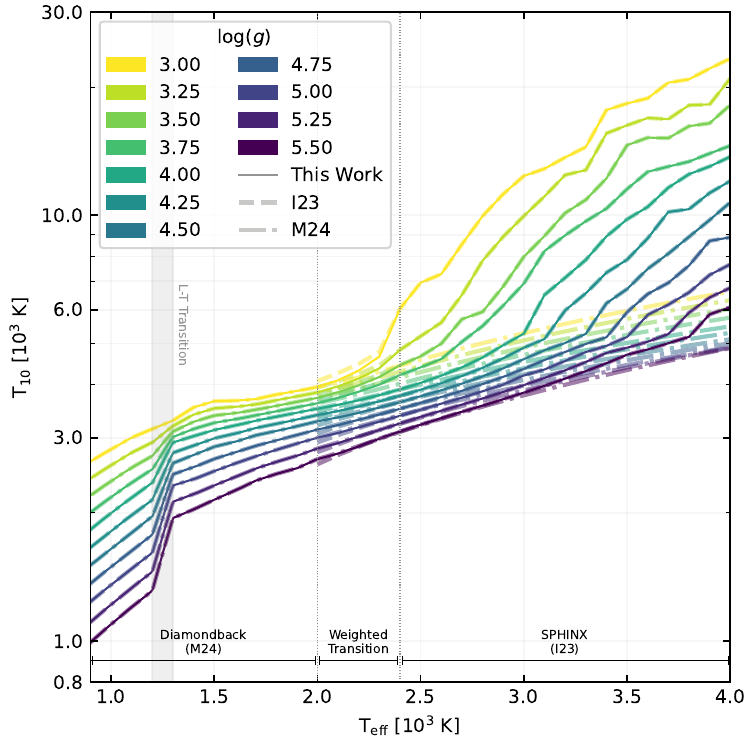}
    \caption{The solar metallicity ([M/H]$=+$0.0) \Teff{}--log($g$)--\Tten{} table used as a boundary condition in our thermal evolution model.  The \Tten{} values are tabulated for the \texttt{Sonora Diamondback} \citepalias{Morely:2024_SonoraDiamondback} and \texttt{SPHINX} \citepalias{Iyer:2023_SphinxMdwarfAtm} atmospheres as indicated in the plot.  The process used to determine the boundary condition in the ``'weighted transition'' region, where these two atmosphere grids overlap in \Teff, is described in \S \ref{sec:methods:t10}.  A structure model with a particular \Tten{} and surface gravity determines its \Teff{} by interpolating within this table.}
    \label{fig:logZ+0.0_T10Table}
\end{figure*}

Figure \ref{fig:PT_Comparison} shows a comparison at \Teff{} of 2400 K and 3000 K between the two codes.  The temperature at $P=$ 10 bar, \Tten, serves as an important boundary condition and is indicated by a horizontal thin dotted gray line.  For reference, the \Tten{} values from \DBK{} are denoted as colored ``X'' markers.  At \Teff{} $=$ 2400 K (left panel), the agreement at log($g$) $=$ 4.5 and 5.5 in the deep atmosphere is excellent.  In the upper atmosphere, however, the \SPX{} profiles are several hundred degrees cooler than those of \DBK.  This is due to the inclusion of silicate (\ce{MgSiO4}, \ce{MgSiO3}), iron, and corundum (\ce{Al2O3}) clouds in \DBK.  In general, clouds increase atmospheric opacity, shortening the mean free path and causing photons that would otherwise escape to heat the atmosphere.  Thus, \DBK{} showed that thicker clouds (lower f$_\mathrm{sed}$) resulted in upper atmospheres that were several hundred degrees hotter than their non-cloudy atmospheres.  The \SPX{} atmospheres used in this work are cloudless, but the cool (\Teff{}$\lesssim$ 2500 K) atmospheres of late M dwarfs, such as TRAPPIST-1, may contain cloud condensates.  The fitting of SPHINX atmospheres with cloud opacities to observations of M dwarfs is currently in progress (Iyer et al. in prep).


We see the deeper atmospheres of \SPX{} start to become significantly hotter at a given temperature at log($g$) $=$ 3.5.  These models at higher $T$ and lower $P$ are in a regime where \texttt{SPHINX}'s range of neutral atomics start to have a major effect on atmospheric opacities, and the opacities of many of these atomics are not included in \DBK{}, such that \DBK{} models underestimate the opacity and are too cool at log($g$) $=$ 3.5.

The right panel shows \texttt{SPHINX} $P$--$T$ profiles at \Teff{} $=$ 3000 K, along with \emph{extrapolations} of the expected \Tten{} values based \DBK{} calculations at lower \Teff{}.  The agreement between \SPX{} and \DBK{} is very good at log($g$) $=$ 5.5, the ``coolest" and densest atmosphere, but these same \DBK{} extrapolations significantly underestimate the calculated \Tten{} at lower gravities, being several thousand degrees off at log($g$) $=$ 3.5.



\subsection{Atmospheric Boundary Condition} \label{sec:methods:t10}

Following previous work in the field, we draw the interior-atmosphere boundary at $P=$ 10 bar \citep[][\DBK]{Lunine:1989_OpacityBDCooling, Saumon&Marley:2008_LTdwarfEvo}.  Other works \citep[][etc.]{Chabrier&Baraffe:1997_LMSEvo, Baraffe:2002_EarlyBDEvo} draw this boundary at a specified optical depth, typically $\tau=$ 100.  These formalisms are a proxy for matching the deep atmosphere's specific entropy to that of the purely convective interior, and are valid as long as the temperature gradient of the atmosphere is adiabatic at the boundary.  The \SPX{} atmospheres are fully convective at $P\geq$ 10 bar, but a small number of low temperature, high surface gravity \DBK{} atmospheres are not yet convective at this pressure.  In these rare cases, the boundary condition for that \DBK{} atmosphere is calculated by continuing the atmosphere's adiabatic temperature gradient from the deep atmosphere  --- where it becomes and stays adiabatic --- upwards to $P=$ 10 bar.  In essence, this computed value of \Tten{} serves as a proxy for specifying the entropy of the atmosphere's deep adiabat.  One could readily pick a higher pressure boundary condition to ensure adiabaticity, but caution must be taken that the boundary radius $R_\mathrm{bound}$ is approximately the photospheric radius $R_\mathrm{phot}$ for the Stefan-Boltzmann equation $L = 4 \pi R_\mathrm{phot}^2 \sigma T_{\mathrm{eff}}^4$ to hold.  Drawing the atmosphere-interior boundary as we describe here ensures that the boundary condition is always a proper proxy for the deep atmosphere and interior specific entropy \citep{Lunine:1989_OpacityBDCooling} while also maintaining $R_\mathrm{bound} \approx R_\mathrm{phot}$.

The boundary condition is then a table of \Tten{} values across atmospheric effective temperature and surface gravity.  The interior model then further interpolates within this table to a \Teff{} value after calculating the interior \Tten{} and surface gravity.  An example \Tten{} boundary condition table for solar metallicity is depicted in Figure \ref{fig:logZ+0.0_T10Table}.  Some of the values in our \Tten{} tables do not match those calculated from the \SPX{} atmospheres as described above.  These \Tten{} values were linearly interpolated over to maintain monotonic behavior in each table, effecting 2.7\%, 4.8\%, and 6.4\% of the metal-rich, solar metallicity, and metal-poor \SPX{} boundary condition points, respectively.  The ``weighted transition'' region in Figure \ref{fig:logZ+0.0_T10Table} is where the \SPX{} and \DBK{} atmospheres overlap in \Teff{} (2000--2400 K).  Here, we smoothly weave these two grids together by sampling both \Tten{} grids between 2000--2400 K at 100 K steps. The weaving of the \DBK{} and \SPX{} \Tten{} grids is calculated as:


\begin{align}
    T_{10} = 
    \begin{cases}
        T_{10,\text{M24}} & T_\mathrm{eff} < 2000\mathrm{K} \\
        \left(\frac{2400-T_\mathrm{eff}}{400}\right)T_{10,\text{M24}}& \\ + \left(\frac{T_\mathrm{eff}-2000}{400}\right) T_{10,\text{I23}}, & 2000\mathrm{K} \leq T_\mathrm{eff} \leq 2400\mathrm{K} \\
        T_{10,\text{I23}}, & T_\mathrm{eff} > 2400\mathrm{K} \\
    \end{cases}
\end{align}


The figure clearly depicts how drastically \Tten{} increases in the high \Teff{} ($\geq$ 2400 K) \SPX{} atmosphere profiles versus the \DBK{} extrapolations.  Within the ``weighted transition'' region between 2000 K and 2400 K --- where there are atmospheric models from both \SPX{} and \DBK{} --- the \SPX{} \Tten{} are typically within 100 K of \DBK, showing excellent agreement.  At low gravities (log($g$) $\lesssim$ 3.25--4, depending on metallicity) the disagreement widens by up to a few thousand K.  However, only a small subset of objects with masses around the deuterium burning limit ($M=$ 0.0125 \Msun{} $=$ 13.1 \Mjup), for a very small range of time (age $\sim$ 1--5 Myr), have sufficiently high \Teff{} and low log($g$) to be in this region.  As \Teff{} rises above 2400 K --- where \DBK{} \Tten{} values are extrapolated --- \SPX{} models become significantly hotter (sometimes upwards of 10$^4$ K) than the \DBK{} extrapolations, due to the inclusion of atomic opacity sources.  Figures \ref{fig:logZ+0.5_T10Table} and \ref{fig:logZ-0.5_T10Table} in the Appendix further illustrate the differences between the \SPX{} and \DBK{} grids for metal-poor ([M/H] $=-0.5$) and metal-rich ([M/H] $=+0.5$), respectively.  We find that \Tten{} and the differences in \Tten{} between \SPX{} and \DBK{} are positively correlated with metallicity, as more molecular and atomic absorption lead to more heating in the deep atmosphere.


\subsection{Thermal Evolution Model} \label{sec:methods:evo}
For this work we use the same structure and evolution model as that of \DBK, where the methods are described in more detail.  The code is an extension of the giant planet evolution code originally presented in \citet{Throgren:2016_GiantPlanetMZ}, modified to include nuclear burning.  The model has 3 essential components:

\begin{enumerate}
    \item \textit{Interior Structure:} \\ The interior structure model calculates the pressure, density, and temperature structure at 1000 mass shells throughout the interior by adding the H/He equation of state (EoS) of \cite{Chabrier:2019_CMS19EoS} with the metal EoS of \citet{Thompson:1990_MetalEoS} via the additive volume law.  The H, He, and metal mass fractions ($X$, $Y$, and $Z$, respectively) used in this work are listed in Table \ref{tab:MassFracs} and are identical to those in \DBK.  They are computed by taking the proto-solar $Z$ values of \citet{Lodders:2009_SolarAbun} and scaling $X$ and $Y$ from the remaining mass fraction budget given the initial \citet{Chabrier:2019_CMS19EoS} values ($X=$ 0.725 and $Y=$ 0.275).  The interior structure model satisfies mass conservation,
    \begin{equation}
        \frac{\partial r}{\partial m} = \frac{1}{4\pi r^2 \rho}
    \end{equation}
    hydrostatic equilibrium, 
    \begin{equation}
        \frac{\partial P}{\partial m} = -\frac{Gm}{4\pi r^4}
    \end{equation}
    and assumes that the object is fully convective and adiabatic (eg. it has no compositional gradients and is isentropic at all times).  The resulting temperature structure allows for the computation of nuclear burning rates, \Teff{} (see \S \ref{sec:methods:atm}), and the luminosity $L = 4 \pi R^2 \sigma T_{\mathrm{eff}}^4$.
    
    \item \textit{Nuclear Burning:} \\ The nuclear burning calculation follows that of \citet{Saumon&Marley:2008_LTdwarfEvo} and takes into account the H, $^2$H (deuterium), and ``effective'' He mass fractions ($X$, $X_\mathrm{D}$, and $Y'$, respectively) in computing the nuclear energy generation rate per unit mass $\epsilon_{\mathrm{nuc}}$.  Because no metal mass fraction is explicitly included in these calculations, both this work and \DBK{} choose an effective He mass fraction, $Y'$, that adds the structure model's metal and He mass fractions such that $Y'=Y+Z$ and $X+Y'=1$ \citep{Chabrier&Baraffe:1997_LMSEvo}.
    
    We note that the initial mass fractions in the nuclear burning code for this work deviate from those of \DBK.  The \DBK{} nuclear burning calculations were initiated with the same mass fractions based on proto-solar values from \citet{Lodders:2009_SolarAbun} ($X=$ 0.7112, $X_\mathrm{D}=$ 2.88$\times$10$^{-5}$, and $Y'=$ 0.2888) for all metallicities.  In this work, the initial $X$ and $Y'$ are still based on those of \citet{Lodders:2009_SolarAbun} but are metallicity dependent, and $X_\mathrm{D}=$ 2.88$\times$10$^{-5}$ for all metallicities.  In reality, the initial deuterium abundance would change with metallicity, assuming a fixed $^2$H/$^1$H ratio.  For the sake of simplicity, and because there are significant uncertainties in early evolution of these objects due to the initial $X_\mathrm{D}$ \citep{Baraffe:2002_EarlyBDEvo}, we choose to fix the initial deuterium abundance for all metallicities.  The $X$, $X_\mathrm{D}$, and $Y'$ used in this work are found in Table \ref{tab:MassFracs}.
    
        
    \item \textit{Time Evolution:} \\ The time step $dt$ is the length of time it takes for the current model to experience a change in entropy $dS$.  Using the luminosity ($L=4\pi R^2\sigma T_\mathrm{eff}^4$), $\epsilon_{\mathrm{nuc}}$, and interior temperature structure calculated above, $dt$ is calculated using Equation \ref{eqn:timeevocalc}.  
    
    \begin{equation}
        \left( L-\int_0^M \epsilon_{\mathrm{nuc}}dm \right)dt = -\int_0^M TdSdm
        \label{eqn:timeevocalc}
    \end{equation}

    This process is repeated for the next time step $t+dt$ until a final age specified by the user is reached.  To ensure the code's solver does not take too large a time step (i.e. too large of changes in density, temperature, etc.) during sensitive evolutionary phases (e.g.  deuterium burning), we update the time evolution code in this work so that $dt$ is limited to 5\% of the current age of the object.  Placing more stringent limits on the maximum time step imperceptibly changed our evolution tracks.  This update fixes an issue in the \DBK{} evolution tracks where deuterium burning ended abruptly and unpredictably, which caused some of the \DBK{} evolution tracks to have moderately lower luminosities, effective temperatures, and radii towards the end of deuterium burning compared to the evolution tracks presented here .

\end{enumerate}

\begin{figure*}[ht]
    \centering
    \includegraphics[width=0.75\linewidth]{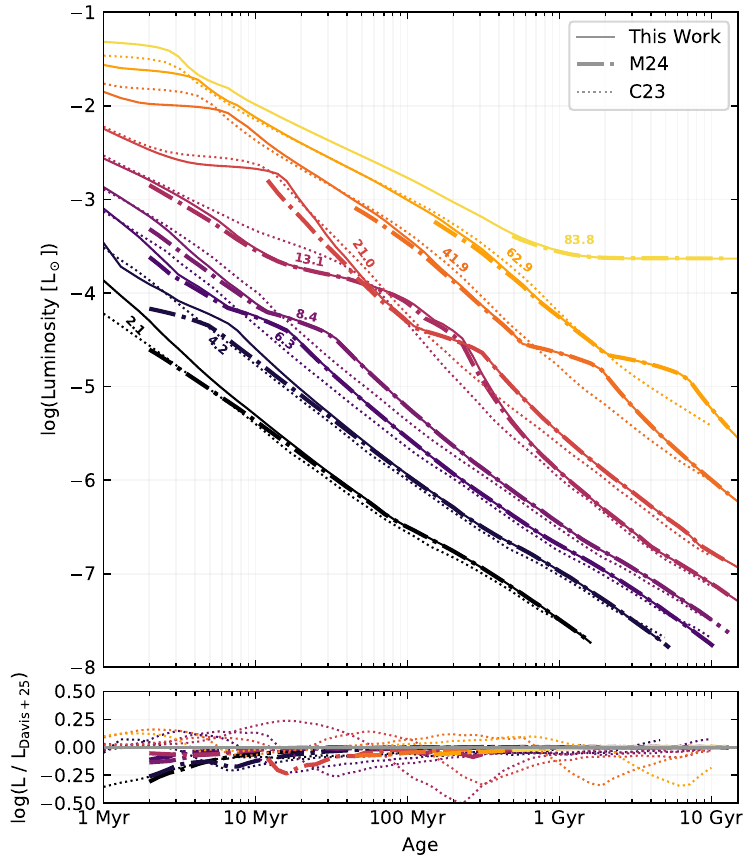}
    \caption{Luminosity evolution tracks of selected models from our work (thin solid lines) compared to the \DBK{} tracks (thick dot-dashed lines) and the recent \CBPD{} models (thin dotted lines).  Models are labeled by their mass in \Mjup.  The bottom panel shows the logarithmic ratios of the \DBK{} and \CBPD{} models to those of this work, (i.e.  a value of 0.25 means the model is $\sim$1.78x as luminous as our models at that age.) Our luminosity tracks are typically within 1\% of the \DBK{} luminosities at late ages ($\gtrsim$ 1 Gyr), but are more luminous at young ages due to higher initial entropy and the inclusion of \SPX{} atmospheric boundary conditions.  As the \CBPD{} models do not account for the L--T transition, they tend to be more luminous before and less luminous after the L--T transition than our models except for the very lowest mass cases, which are less luminous at all young ages and more luminous beyond several Gyr.}
    \label{fig:ModelComp_L-vs-t}
\end{figure*}

\section{Results}




\begin{figure*}[ht]
    \centering
    \includegraphics[width=0.75\linewidth]{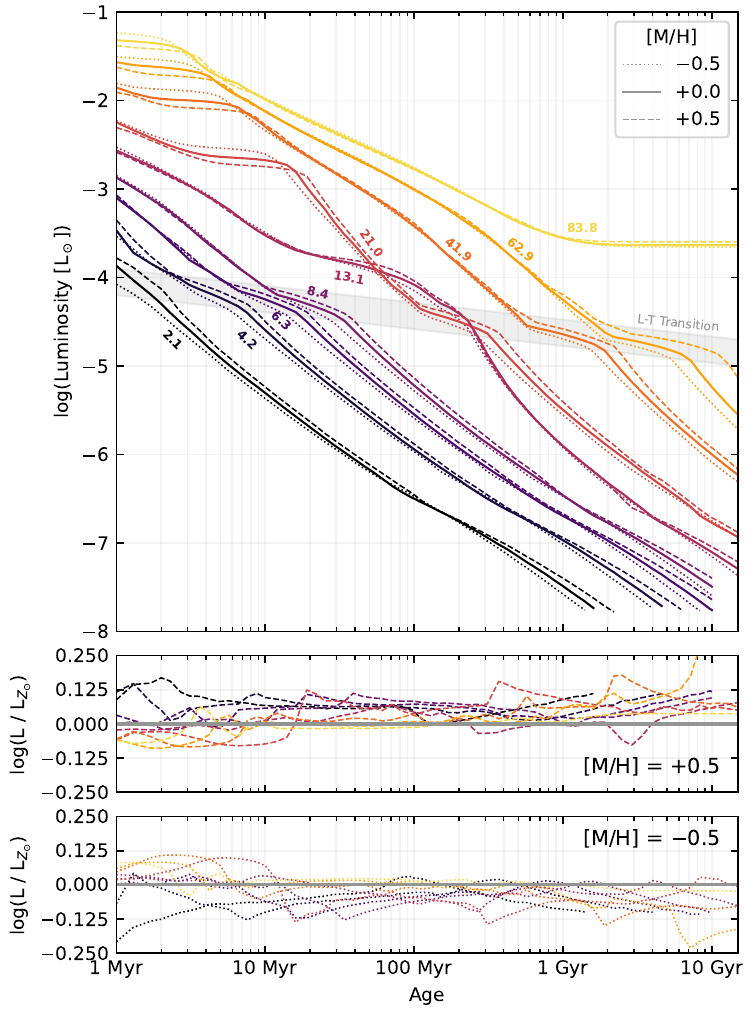}
    \caption{Luminosity evolution tracks (top panel) of selected solar metallicity ([M/H]$=+$0.0, thick solid lines), metal-rich ([M/H]$=+$0.5, thin dashed lines) and metal-poor ([M/H] $=-$0.5, thin dotted lines).  Models are labeled by their mass in \Mjup.  We show the logarithmic ratios of the metal-rich (middle panel) and metal-poor (bottom panel) models to the solar metallicity models, (i.e.  a value of 0.125 means the model is $\sim$1.33x as luminous as its solar metallicity counterpart at that age.) Higher mass objects are less luminous the more metal rich they are until $\sim$10 Myr, where they become more luminous.}
    \label{fig:AllZ_L-vs-t}
\end{figure*}

\begin{table*}[]
\centering
\caption{Mass fractions used in our models.  In the structure code, the solar metallicity $Z$ value of 0.0153 is that of \citet{Lodders:2009_SolarAbun}, while the values of $X$ and $Y$ are scaled down from the pure hydrogen-helium mixture of \citet{Chabrier:2019_CMS19EoS} ($X=$ 0.725 and $Y=$ 0.275) using the additive volume law such that $X+Y+Z=1$.  In contrast, the nuclear burning code uses the proto-solar \citet{Lodders:2009_SolarAbun}  $X$, $Y$, and $Z$ mass fractions, and adds the metal mass fraction $Z$ to the helium mass fraction $Y$ such that $Y'=Y+Z$ and $X+Y'=1$.  We specify a deuterium mass fraction $X_\mathrm{D}=2.88\times10^{-5}$ in the nuclear burning code for all metallicities.}
\begin{tabular}{ccc} \hline\hline
[M/H] & Structure Code ($X+Y+Z=1$) & Nuclear Code ($X+Y'=1$) \\
        & $X$ ||| $Y$ ||| $Z$      & $X$ ||| $Y'$ ||| $X_\mathrm{D}$ \\ \hline
$+$0.5  & 0.6914 | 0.2623 | 0.0463 & 0.6888 | 0.3112 | 2.88$\times$10$^{-5}$     \\
$+$0.0  & 0.7139 | 0.2708 | 0.0153 & 0.7112 | 0.2888 | 2.88$\times$10$^{-5}$     \\
$-$0.5  & 0.7214 | 0.2737 | 0.0049 & 0.7187 | 0.2813 | 2.88$\times$10$^{-5}$     \\ \hline
\end{tabular}
\label{tab:MassFracs}
\end{table*}

\begin{figure}[ht]
    \centering
    \includegraphics[width=0.825\linewidth]{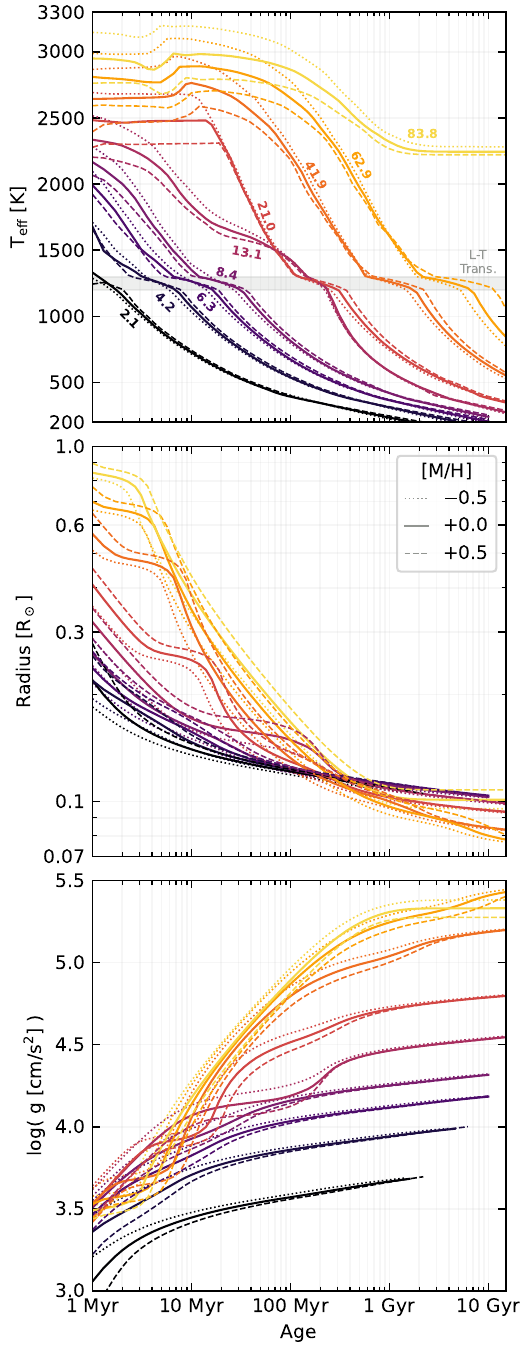}
    \caption{Evolution tracks of \Teff{} (top panel), radius (middle panel), and surface gravity (bottom panel) of selected solar metallicity (thick solid lines), metal-rich (thin dashed lines) and metal-poor (thin dotted lines) model objects.  Models are labeled by their mass in \Mjup.  In general, the objects tend to be hotter, smaller, and of higher surface gravity as metallicity decreases, with these patterns becoming more apparent at higher masses and younger ages.  Most of the modeled properties here become less sensitive to metallicity as these objects age, but objects that are close to or above the hydrogen-burning limit ($M\gtrsim$ 0.06\Msun{} $=$ 62.9 \Mjup) retain differences due to metallicity well after a few Gyr.}
    \label{fig:AllZ_TRG-vs-t}
\end{figure}


\subsection{Evolution Tracks} \label{sec:results:evotracks}

\begin{figure*}[ht]
    \centering
    \includegraphics[width=0.75\linewidth]{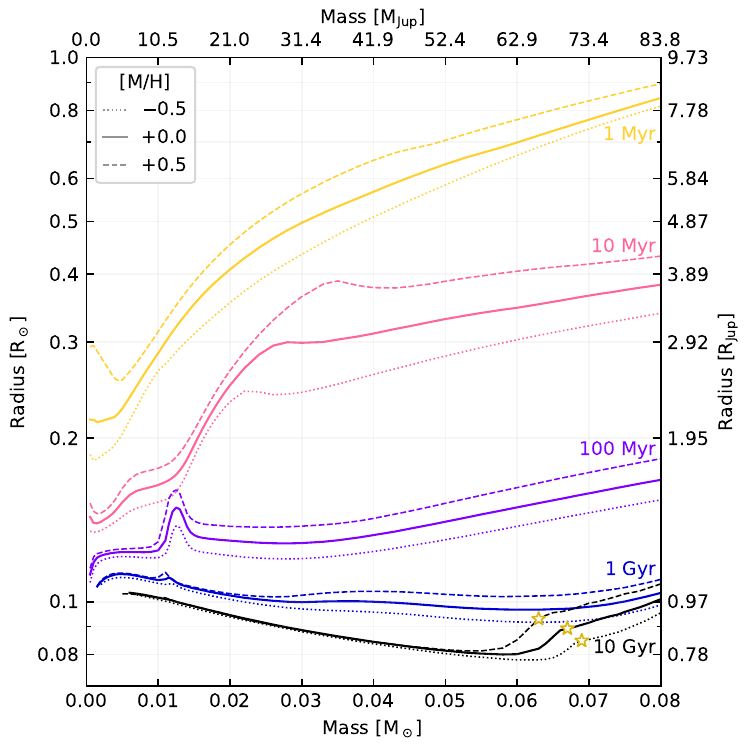}
    \caption{Radii of our models as a function of mass for metal-poor (dotted lines), solar metallicity (solid lines), and metal-rich (dashed lines) at various ages.  Higher mass objects contract more than lower mass objects between 1 Myr and 10 Gyr.  Higher metallicity is correlated with larger radii at a given age, owing to increased opacities which slow contraction.  The bump in the 100 Myr contours around 0.010--0.015 \Msun{} (10.5--15.7 \Mjup) is the result of deuterium-burning, which temporarily halts contraction.  The 10 Gyr contours shows that higher metallicity leads to a lower hydrogen-burning minimum mass (HBMM), as indicated by the gold stars between 0.06--0.07 \Msun{} (62.9--73.4 \Mjup).  The HBMM for each metallicity is identical to the \DBK{} ``hybrid'' models.}
    \label{fig:AllZ_M-vs-R}
\end{figure*}

\subsubsection{Luminosity} \label{sec:results:evotracks:L}

Figure \ref{fig:ModelComp_L-vs-t} shows our calculations of luminosity evolution and compares them to \DBK.  Our tracks agree extremely well with \DBK{} (typically within 1\%) at later ages.  Because our boundary condition is identical to \DBK{} below 2000K (see above text and Fig. \ref{fig:logZ+0.0_T10Table}), all our simulated objects align well with the \DBK{} tracks at $\gtrsim$1 Gyr.  However, at young ages ($\lesssim$10--100 Myr), our tracks become more luminous than \DBK{} because 1) we assume a higher initial entropy for the arbitrarily hot start for each model \citep[see][]{Marley:2007_YoungGiantLum}, and 2) the updated \SPX{} boundary conditions limit how quickly these objects can cool, altering the cooling history at all younger ages.  For objects under $\sim$0.011 \Msun{} (11.5 \Mjup), the difference is entirely due to higher initial entropy as they are never hot enough to be impacted by the updated boundary conditions.  At intermediate masses (0.011 \Msun $\lesssim M \lesssim$ 0.05 \Msun, 11.5 \Mjup $\lesssim M \lesssim$ 52.4 \Mjup), effects from initial entropy are lost well before 1 Myr and the \SPX{} boundary conditions become the dominant factor.  Above 0.05 \Msun (52.4 \Mjup), the \SPX{} boundary condition is the primary factor in setting cooling timescales while initial entropy plays a small but non-negligible part.

The figure depicts several notable evolutionary stages of the objects.  Deuterium-burning is the cause of the first temporary ``plateau'' in luminosity for objects above the deuterium burning limit (the model plotted just above this limit is 0.0125 \Msun{} $=$ 13.1 \Mjup).  This takes place in the first few Myr for the most massive objects and at progressively later ages for less massive objects.  The near-plateaus around $L\sim10^{-4}$--$10^{-5} L_\odot$ are due to the L--T transition, which is modeled as the result of a hotter photosphere being exposed as the clouds break up in the atmosphere \citep[][\DBK]{Saumon&Marley:2008_LTdwarfEvo}.  For the highest mass objects ($M\gtrsim$ 0.067 \Msun{} $=$ 70.2 \Mjup, depending on metallicity), the final plateau continues out to 15 Gyr, indicating that these objects have achieved sufficient energy from H-burning to halt their collapse and thus entered the main sequence. The effect of metallicity on the HBMM and a comparison of this work to \DBK{} are discussed in \S \ref{sec:results:MvsR}.


%

Similarly, Figure \ref{fig:AllZ_L-vs-t} shows the luminosity evolution tracks of our models with different metallicities.  The middle and bottom panels show the logarithm of the ratio between the metal-rich ([M/H] $=+$0.5) and metal-poor ([M/H] $=-$0.5) models, respectively, to the solar metallicity models.  After 10--20 Myr, metal enrichment leads to more luminous objects, with objects below 0.06\Msun{} (62.9 \Mjup) retaining this relationship back to 1 Myr.  Metal-rich objects with $M\geq$ 0.06 \Msun{} $=$ 62.9 \Mjup{} are slightly less luminous ($\lesssim$ 0.02 dex, $\lesssim$ 5\%) than solar metallicity objects, with the inverse being true for metal-poor objects.  Higher metallicities cause higher atmospheric opacities that slow cooling and result in typically brighter objects, except for objects of mass $M\geq$ 0.02 \Msun{} $=$ 21.0 \Mjup, which have lower luminosities around deuterium burning.  Higher metallicity also pushes deuterium burning to later ages for objects above 0.0125\Msun{} (13.1\Mjup).

\subsubsection{Effective Temperature} \label{sec:results:evotracks:Teff}

Figure \ref{fig:AllZ_TRG-vs-t} shows the effective temperature, radius, and surface gravity evolution tracks for the same sample of models above.  The top panel shows that the effective temperature evolution tracks looks similar to the luminosity tracks of Fig. \ref{fig:AllZ_L-vs-t} except at ages less than a few Myr, where objects above the deuterium burning limit have their effective temperature and radii stabilized against contraction during deuterium burning.  Once deuterium burning has ceased, the radius of these more massive objects drops precipitously while effective temperature changes little, indicative of these objects being on the Hayashi track \citep{Hayashi:1961_HayashiTrack} and that their interiors are not yet dominated by degeneracy pressure.  Additionally, higher metallicity objects are typically cooler before and become hotter after the L--T transition at \Teff{}$=$1300 K, as the blanketing effect of clouds, and therefore the strength of the L--T transition, grows stronger with more metals in the atmosphere.

\subsubsection{Radius} \label{sec:results:evotracks:R}

The middle panel shows that higher metallicity objects are consistently larger in radius, as higher opacities increase the cooling timescale.  While objects below the hydrogen-burning minimum mass (HBMM) eventually cool enough that opacity effects begin to subside and radii become less dependent on metallicity, objects above this limit stop cooling and contracting at temperatures where the atomic and molecular opacities still matter greatly, preserving the metallicity-dependence of their radii.  

\subsubsection{Surface Gravity} \label{sec:results:evotracks:logg}

The bottom panel shows that the majority of objects have a surface gravity of log($g$) $\sim$ 3.5 around 1 Myr, but quickly ``fan out" with age, covering a larger range that becomes increasingly dependent on mass.  For example, objects more massive than $\gtrsim$ 42 \Mjup{} quickly rise to log($g$) $\sim$ 4.5 by 20--30 Myr, but objects near the deuterium burning limit ($\sim$12 \Mjup) do not reach this same surface gravity until they are $\sim$2--3 Gyr old. We note that some of the smallest mass objects have log($g$) $<$ 3 at extremely young ages ($\lesssim$ 10 Myr), which is outside our atmospheric boundary condition grid (see Fig. \ref{fig:logZ+0.0_T10Table}). The extrapolation in this region of \Teff{}--log($g$) space affects the models at a mass of 0.0025 \Msun{} (2.6 \Mjup) at 1 Myr, and the lowest mass case (0.0005 \Msun{}, 0.5 \Mjup) out to 96 Myr.

\begin{figure*}[ht]
    \centering
    \includegraphics[width=0.75\linewidth]{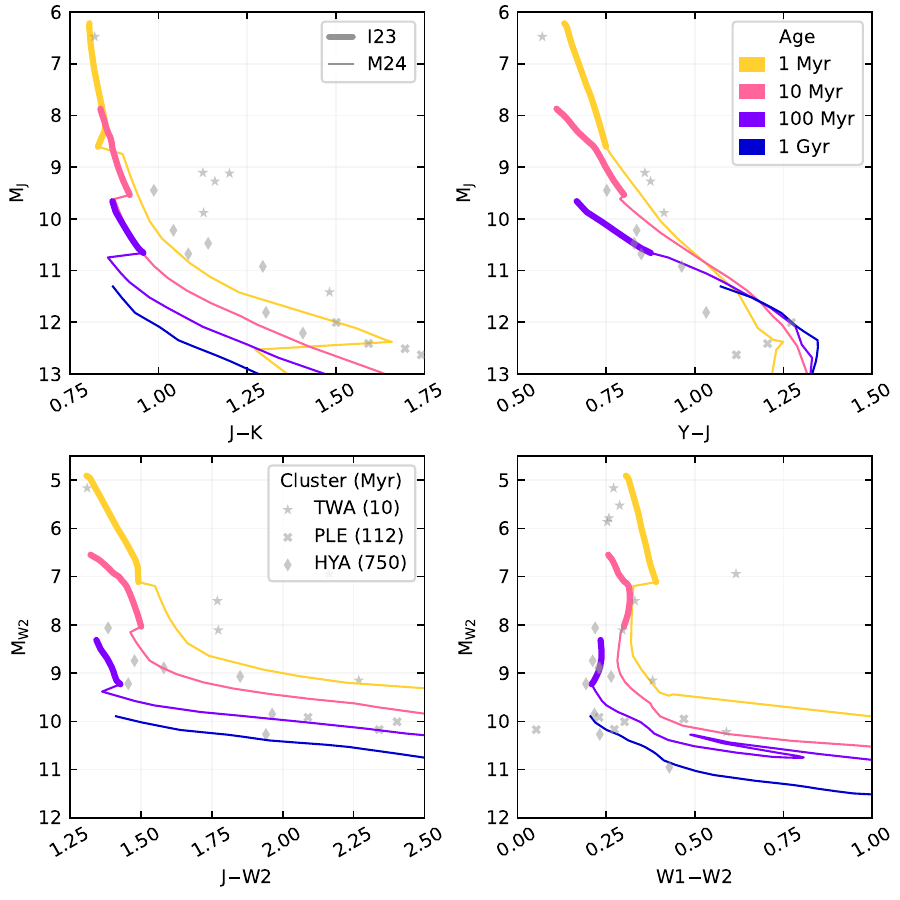}
    \caption{Color-magnitude diagrams and isochrones in various filters.  Thick lines refer to portions of isochrones that come from \SPX{} synthetic photometry, while thin lines refer to parts of isochrones from \DBK{} synthetic photometry. We include empirical photometry of the TW Hydrae (TWA, grey stars), Pleiades (PLE, grey crosses), and the Hyades (HYA, grey diamonds) young clusters via the UltracoolSheet \citep{Best:2024_UltracoolSheetZenodo}.}
    \label{fig:CMDs_isochrones}
\end{figure*}


\subsection{Mass-Radius Relationship} \label{sec:results:MvsR}

Figure \ref{fig:AllZ_M-vs-R} depicts the mass-radius relationship for our models as a function of age.  It is equivalent to turning the middle panel of Fig. \ref{fig:AllZ_TRG-vs-t} ninety degrees clockwise into the page.  Increased atmospheric opacity from higher metallicity slows cooling, leading to slower contraction and a larger radius at a given age.  As described above, once objects at old ages ($\gtrsim$ 1 Gyr) have sufficiently cooled, contracted, and their interiors have become high in density and supported by electron degeneracy pressure, their radii become less dependent on metallicity .

The ``shoulder" in the 10 Myr (pink) contour at masses 0.02--0.035 \Msun{} (21.0--36.7 \Mjup) is due to deuterium-burning.  Later deuterium-burning at masses 0.01--0.015 \Msun{} (10.5--15.7 \Mjup) leads to the pronounced radius bump at 100 Myr (purple).  Remnants of this altering of thermal evolution can still be seen at 1 Gyr (dark blue).  


The 10 Gyr contours show how the HBMM moves to lower masses as metallicity increases.  Higher metallicity introduces more absorbers and opacity into the atmosphere, which a) steepens the temperature gradient, leading to a higher central temperature, which increases the thermonuclear reaction rate, and b) limits the power that can be radiated away by the objects, thus reducing the luminosity and thermonuclear reaction rate necessary to achieve the fusion-gravity balance that is the hallmark of a star \citep[as discussed extensively in][]{Burrows:2001_BDReview}.  This results in the 10 Gyr contour's ``shelf'' moving to lower masses and larger radii with increased metallicity.  We find our HBMM and deuterium-burning limit are identical to the ``hybrid'' models of \DBK{} for all metallicities: $+$0.5: 0.063 \Msun{} (66.0 \Mjup), $+$0.0: 0.067 \Msun{} (70.2 \Mjup), and $-$0.5: 0.069 \Msun{} (72.3 \Mjup).  By the time all three of these models reach the main sequence, \Teff{} has fallen well below 2000 K, where our boundary condition is identical to \DBK{} (see \S \ref{sec:methods:t10} and Fig. \ref{fig:logZ+0.0_T10Table}.)

\subsection{Isochrones and Color-Magnitude Diagrams}

In Figure \ref{fig:CMDs_isochrones}, we present solar metallicity isochrones at several different ages.  These isochrones were computed in a simple framework: first, we tabulate synthetic photometry --- publicly available at the Zenodo repository \footnote{https://doi.org/10.5281/zenodo.15611936} --- across the grid of effective temperature and surface gravity from \SPX{} and \DBK{}. We choose to compute photometry using \SPX{} spectra at \Teff{} $\geq$ 2400 K, and \DBK{} spectra when \Teff{} $<$ 2400 K, using the nearest radius in our evolution model grid to the surface gravity and metallicity of each spectrum.  Then, we calculate the effective temperature and surface gravity for each mass in our evolution model grid at the specified age and metallicity, and magnitudes are linearly interpolated within the synthetic photometry grid to these calculated \Teff{} and log($g$) values.

We compare our isochrones to empirical photometry of young M and L type (M4--L9) published in the Ultracool sheet \citep{Dupuy:2012_UltracoolSheetI, Dupuy:2013_UltracoolSheetII, Liu:2016_UltracoolSheetIII, Best:2018_UltracoolSheetIV, Best:2021_UltracoolSheetV, Best:2024_UltracoolSheetZenodo}.  The objects are members of the young associations: members of TW Hydrae (TWA) are marked with grey stars, Pleiades (PLE) with grey crosses, and Hyades (HYA) with grey diamonds.  In each panel, we only plot objects with $\leq$20\% error in parallax and $\leq$10\% error in both magnitudes for that panel, and we remove binaries according to the \texttt{multiplesystem\_unresolved\_in\_this\_table} key.

Individual isochrones typically move towards larger magnitudes and colors as mass decreases, indicative of objects of lower mass being fainter and redder.  Additionally, some isochrones have nearly lateral jumps blue-ward by $\sim$0.1--0.2 magnitudes in color when transitioning from \SPX{} spectra (thick lines) and \DBK{} spectra (thin lines), and are due to modest differences between spectra.  The 1 Myr isochrone shows a sharp jump blue-ward in J$-$K (top left panel) near a J band magnitude of 12.5, which coincides with the L--T transition at \Teff{} $=$ 1300 K.  \DBK{}'s Figure 8 shows that cloud-free spectra of \Teff{} $\approx$ 1300 K are significantly bluer in J-K than their cloudy counterparts.  Thus, as the ``hybrid'' \DBK{} models used in this work transition from cloudy (f$_\mathrm{sed}=$ 2) above \Teff{} $=$ 1300 K to cloud-free models below \Teff{} $=$ 1300 K, the spectra and photometry make a sharp jump blue-ward.  The 100 Myr isochrone also shows a sharp blue-ward jump in W1$-$W2 over a short range of masses (0.010--0.015 \Msun, 10.5--15.7 \Mjup).  The deuterium burning stages of these extremely low mass brown dwarfs end near 100 Myr such that their \Teff{} are still 100--200 K hotter (see top panel of Fig. \ref{fig:AllZ_TRG-vs-t}), and therefore are up to 0.3 magnitudes bluer in W1$-$W2 and 0.5 magnitudes brighter in W2, than objects 50--100\% more massive. This phenomena is present in the other filters, but at magnitudes and colors not visible in the isochrones we plot here.

\section{Discussion} \label{sec:discussion}
A theory for the cooling history of brown dwarfs, across six orders of magnitude in luminosity and two orders of magnitude in gravity, presents a host of challenges.  These challenges are likely most significant at young ages, given uncertainties in formation mechanisms.  Many of the main uncertainties in brown dwarf formation --- and how they affect their early evolution --- were explored in detail in \citet{Baraffe:2002_EarlyBDEvo} and \citet{Marley:2007_YoungGiantLum} and are still with us two decades later.

The accretion and proto-stellar collapse set the very earliest physical conditions of massive brown dwarfs, imparting additional material (including deuterium) and/or thermal energy \citep[][and references therewithin]{Hartmann:1997_DiskAccrEvo}.  Despite their importance, these processes are obscured by dust and gas, with objects historically being observable only after these processes have taken place \citep{Baraffe:2002_EarlyBDEvo}.  These processes are an important source of uncertainty in the choice of initial conditions for 1D thermal evolution models.  


It seems likely that further improvements in 1D evolution models will come from a better understanding of  atmospheres.  Fitting spectra over a broad range surface gravities, at \Teff{} values above, within, and below where cloud opacity dominates, will allow for a better understanding of temperature structures. 



A recently published suite of models \citep[][hereafter \CBPD]{Chabrier:2023_CD21newHBMM} also extended back to 1 Myr and primarily focused on how a new H/He EoS \citep{Chabrier:2021_CD21EoS} resulted in faster cooling timescales of brown dwarfs, larger masses for a given age and temperature, and a larger HBMM.  We compared the luminosity evolution tracks of this study to our own in Figure \ref{fig:ModelComp_L-vs-t}.  Like this study, \CBPD{} found high mass objects to be on Hayashi tracks \citep{Hayashi:1961_HayashiTrack}, where effective temperature is relatively stable while luminosity drops, at young ages ($\lesssim$10 Myr).  However, the models presented here show significant differences in the predicted luminosity around the L--T transition and moderately different behavior around deuterium burning.  The differences in luminosity surrounding the L--T transition are simply because our atmospheric boundary conditions account for opacities resulting from cloud formation at \Teff{}$>1300$ K, and with loss of clouds at  \Teff{}$<1300$ K, while those of \citet{Phillips:2020_ATMO2020evo} do not treat clouds.  

The deuterium burning differences are more complicated: for example, the 0.0125\Msun{} (13.1 \Mjup) \CBPD{} model is $\sim$1.8$\times$ as bright as our model at approximately 15 Myr, but falls to just $\sim$0.32$\times$ by $\sim$200 Myr.  Our deuterium burning phases start and end later and occur at lower luminosities than \CBPD{}.  While it is possible that these discrepancies in deuterium burning are due to differences in initial conditions (eg. $[D]_0$, see above text) or the EoS, we believe it is more likely a consequence of different atmospheric boundary conditions.  Though the evolution models presented by \CBPD{} use an atmosphere grid with log($g$) $=$ 2.5--5.5 and \Teff{} $\leq$ 3000 K, the atmospheres are most valid for \Teff{} $\leq$ 2000 K due to missing high-temperature opacities \citet{Phillips:2020_ATMO2020evo}.  Indeed, \CBPD{} suggest that remaining discrepancies in brown dwarf models to observations are likely to be found in atmospheric boundary condition uncertainties.

Additionally, using the \citet{Chabrier:2021_CD21EoS} EoS in our models indiscernibly affected luminosity for objects above the planet-brown dwarf boundary and produced slightly more compact, less luminous objects at masses below this boundary.  We therefore suggest that the majority of the discrepancies between this work and \CBPD{} are due to various limitations and differences between our atmospheric boundary conditions, rather than slight differences in the initial deuterium abundance or equation of state.

\section{Conclusion}

We present \texttt{Sonora Red Diamondback}, a grid of brown dwarf evolution models based on \citet{Morely:2024_SonoraDiamondback} spanning ages of 1 Myr--15 Gyr, with younger ages enabled by the inclusion of hotter M-type (\Teff{} $=$ 2000--4000 K) atmospheric boundary conditions from \texttt{SPHINX} \citep{Iyer:2023_SphinxMdwarfAtm}.  These models supersede the \citet{Morely:2024_SonoraDiamondback} evolution tracks due to improvements in treating evolution at early ages.  We predict significantly different evolution tracks at young ages (1--100 Myr) compared to what would be found from extrapolations of evolution models meant for older, more compact, and cooler brown dwarfs.  These new models include the appearance of Hayashi tracks and an updated treatment of deuterium burning for objects above the planet-brown dwarf boundary ($\sim$0.0115 \Msun{} $=$ 12 \Mjup), and hydrogen burning, main sequence phases for the highest mass objects.  We find significantly higher boundary temperatures for \Teff{} $\geq$ 2400 K, caused by increased atmospheric opacities due to atoms that are significant at these higher temperatures.  We also find that metallicity has a significant impact on the early evolution of these objects through the substellar mass range. The evolution tracks and synthetic photometry are made available for use by the community at the following Zenodo DOI: \href{https://doi.org/10.5281/zenodo.15611936}{10.5281/zenodo.15611936}.

\section*{Acknowledgments}
CED acknowledges support from the NSF Graduate Research
Fellowship Program.  JJF acknowledges the support of NSF AAG Award, 2009592.  This work has benefited from ``The UltracoolSheet'' at \url{http://bit.ly/UltracoolSheet}, maintained by Will Best, Trent Dupuy, Michael Liu, Rob Siverd, and Zhoujian Zhang and developed from compilations by \citet{Dupuy:2012_UltracoolSheetI, Dupuy:2013_UltracoolSheetII, Liu:2016_UltracoolSheetIII, Best:2018_UltracoolSheetIV, Best:2021_UltracoolSheetV, Sanghi:2023_UltracoolSheetVI, Schneider:2023_UltracoolSheetVII}.  This work benefited from the 2025 Exoplanet Summer Program in the Other Worlds Laboratory (OWL) at the University of California, Santa Cruz, a program funded by the Heising-Simons Foundation and NASA.  This research has made use of NASA’s Astrophysics Data System Bibliographic Services.

\bibliography{chevdavi}
\bibliographystyle{aasjournal}

\appendix

\section{Additional Boundary Condition Tables}

\begin{figure*}[ht]
    \centering
    \includegraphics[width=0.75\linewidth]{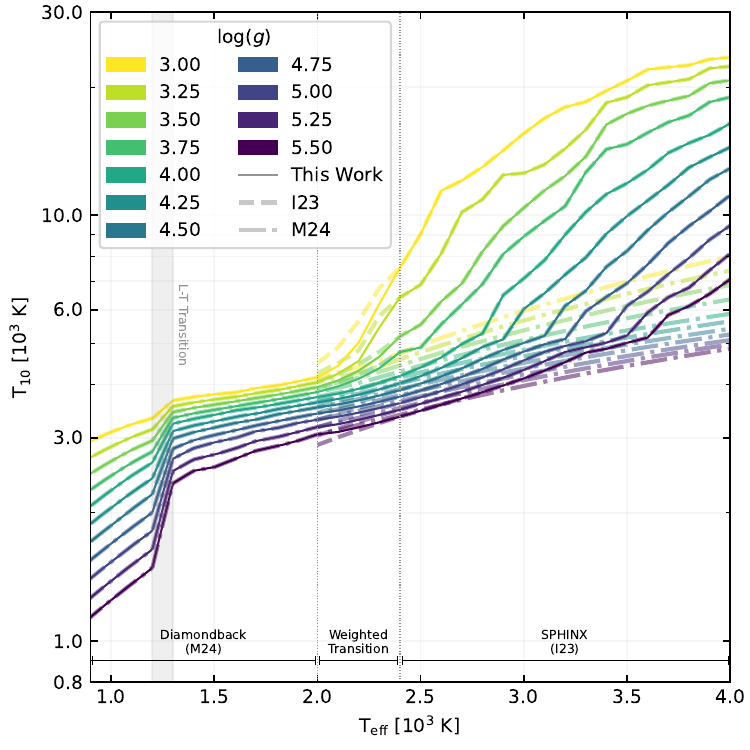}
    \caption{The metal-rich ([M/H]$=+$0.5) \Teff{}--log($g$)--\Tten{} table used as a boundary condition in our thermal evolution model.  The \Tten{} values are tabulated for the \texttt{Sonora Diamondback} \citepalias{Morely:2024_SonoraDiamondback} and \texttt{SPHINX} \citepalias{Iyer:2023_SphinxMdwarfAtm} atmospheres as indicated in the plot.  The process used to determine the boundary condition in the ``'weighted transition'' region, where these two atmosphere grids overlap in \Teff, is described in \S \ref{sec:methods:t10}.  A structure model with a particular \Tten{} and surface gravity determines its \Teff{} by interpolating within this table.}
    \label{fig:logZ+0.5_T10Table}
\end{figure*}

\begin{figure*}[ht]
    \centering
    \includegraphics[width=0.75\linewidth]{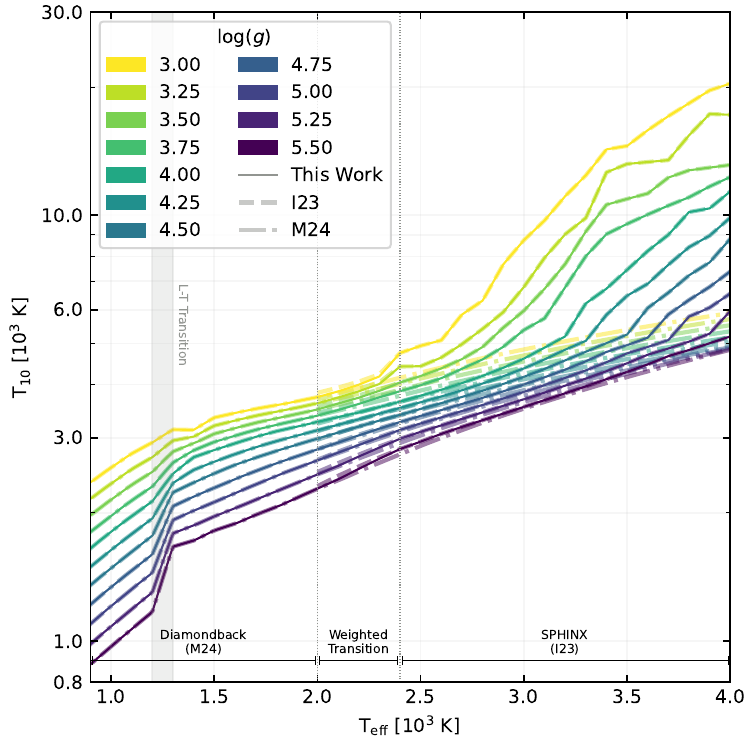}
    \caption{The metal-poor ([M/H]$=-$0.5) \Teff{}--log($g$)--\Tten{} table used as a boundary condition in our thermal evolution model.  The \Tten{} values are tabulated for the \texttt{Sonora Diamondback} \citepalias{Morely:2024_SonoraDiamondback} and \texttt{SPHINX} \citepalias{Iyer:2023_SphinxMdwarfAtm} atmospheres as indicated in the plot.  The process used to determine the boundary condition in the ``'weighted transition'' region, where these two atmosphere grids overlap in \Teff, is described in \S \ref{sec:methods:t10}.  A structure model with a particular \Tten{} and surface gravity determines its \Teff{} by interpolating within this table.}
    \label{fig:logZ-0.5_T10Table}
\end{figure*}

\end{document}